\documentclass[final,12pt]{elsarticle}

\usepackage{graphicx}

\usepackage{amssymb}
\usepackage{amsmath}
\usepackage{color}
\usepackage{comment}

\usepackage{tikz}
\usetikzlibrary{shapes.geometric,arrows}
\usetikzlibrary[arrows.meta,bending]
\usetikzlibrary{positioning}
\tikzstyle{line} = [draw,-latex']

\usepackage[]{todonotes}

\newcommand{\michaeltodo}[2]{\todo[inline,backgroundcolor=orange!20!white]{\color{blue}{{\bf \it  #1}: {\bf Michael:} #2}}}

\definecolor{darkblue}{rgb}{0,0,.6}
\definecolor{darkred}{rgb}{.6,0,0}
\definecolor{darkgreen}{rgb}{0,.6,0}
\definecolor{red}{rgb}{.98,0,0}

\def\ssmall{\fontsize{8pt}{2pt}\selectfont}

\usepackage{fancybox}

\PassOptionsToPackage{normalem}{ulem}
\usepackage{ulem}
\usepackage{subfig}
\usepackage{multirow}
\usepackage{multicol}

\usepackage{colortbl}
\definecolor{lightgray}{gray}{0.8}

\newcounter{bla}

 \usepackage{hyperref}
\usepackage{listings}
\lstnewenvironment{cpplisting}[1][]
  {\lstset{language=C++,numbers=right,#1}}{}

\lstnewenvironment{bashlisting}[1][]
  {\lstset{language=bash,numbers=right,#1}}{}

\lstnewenvironment{outputlisting}[1][]
  {\lstset{language=html,numbers=right,#1}}{}

\journal{Computer Physics Communications}

\begin{document}

\newcommand{\ME}{\begin{tt}Maxent\end{tt}}

\begin{frontmatter}

\title{TorchAmi: Generalized CPU/GPU Implementation of Algorithmic Matsubara Integration}

\author[a,b]{M. D. Burke}
\author[a]{J. P. F. LeBlanc\corref{author}}
\cortext[author] {Corresponding author.\\\textit{E-mail address:} jleblanc@mun.ca}
\address[a]{Department of Physics and Physical Oceanography, Memorial University of Newfoundland, St. John's, Newfoundland \& Labrador, A1B 3X7, Canada}
\address[b]{Department of Physics and Astronomy, University of Waterloo, Waterloo, Ontario, N2L 3G1, Canada}

\begin{abstract}
We present \texttt{torchami}, an advanced implementation of algorithmic Matsubara integration (AMI) that utilizes \texttt{pytorch} as a backend to provide easy parallelization and GPU support.  AMI is a tool for analytically resolving the sequence of nested Matsubara integrals that arise in virtually all Feynman perturbative expansions.  In this implementation we present a new AMI algorithm that creates a more natural symbolic representation of the Feynman integrands. In addition, we include peripheral tools that allow for import and labelling of simple graph structures and conversion to \texttt{torchami} input.  The code is written in c++ with python bindings provided. 
\end{abstract}


\end{frontmatter}



{\bf{PROGRAM SUMMARY}}

\begin{small}
\noindent
{\em Program Title:}  torchami                                        \\
{\em CPC Library link to program files:} \\
{\em Developer's repository link:} \url{https://github.com/mdburke11/torchami} \\
{\em Code Ocean capsule:} (to be added by Technical Editor)\\
{\em Licensing provisions:} GPLv3\\
{\em Programming language:}  \texttt{C++, python}                                 \\
{\em Nature of problem:}\\
Feynman diagrams are pictorial representations of perturbative expansions often formulated in the imaginary frequency/time axis and involve a high-dimensional sequence of nested integral over spatial and temporal degrees of freedom. \\
{\em Solution method:}\\
 \texttt{torchami} provides a framework to symbolically generate and store the analytic solution to the temporal Matsubara sums through repeated application of multipole residue theorems.  The solutions are stored using a tree structure for arbitrary products and sums of Fermi/Bose functions, and the evaluation functions provide both CPU and GPU support with automatic parallelization for batch sampling problems. \\
{\em Additional comments including restrictions and unusual features:}\\
Requires \texttt{C++17} standard, the boost graph library, as well as \texttt{pytorch}. 
   \\

\end{small}

\section{Introduction}
Despite the array of numerical algorithms for tackling many-body problems, few methods are able to access real frequency results for correlated observables.  The development of numerical algorithms therefore remains an active area of study in condensed matter physics.  To this end, a central tool for studying both finite (discrete) and infinite (continuous) systems is the interaction expansion that gives rise to Feynman diagrams.  These diagrams are pictorial representations of physical interaction processes.  These diagrams translate to mathematical expressions that when evaluated provide probability amplitudes for the events they depict.  Quantum field theory dictates that all possible events must be sampled and this means that all internal degrees of freedom must be  summed to obtain the total probability associated with a particular diagram. 

Hence the challenge of many-body perturbation theory collapses to the evaluation of high-dimensional integrals which in general is challenging.  A perk of such a scheme is that the rules for evaluating diagrams remain largely consistent regardless of the specific observable being studied, and thus all problems look similar.  Another advantage is the ability to study both zero and finite temperature properties, though the expansions are convergent only at high-temperatures. 

Nevertheless, integrating the high dimensional space of internal degrees of freedom for arbitrary Feynman diagrams is conceptually straightforward, and has been the topic of a wide range of numerical approaches including diagrammatic monte carlo, determinental quantum monte carlo, and other methods based on the interaction expansion.\cite{vanhoucke,rossi2017determinant}  In all cases, the diagram evaluation includes temporal (times or frequency) and spatial (position or momentum) integrals which are typically treated at the same level and sampled stochastically.  The exception to this approach is algorithmic Matsubara integration (AMI), introduced in Ref.~\cite{AMI}, that has been successfully applied to a number of problems including the uniform electron gas, and the 2D Hubbard model.\cite{mcniven:2021,mcniven:2022,AMI:spin,GIT,igor:spectral,leblanc:2022,taheri:2020, farid:2023,symdet} The name and concept for AMI is borrowed from algorithmic or automatic differentiation.\cite{auto_diff}  As such, AMI is an algorithmic scheme to symbolically generate the analytic result of a subset of the integration space of Feynman diagrams.  Specifically, the evaluation of temporal integrals when formulated in frequency-space can be handled analytically via repeated applications of Cauchy residue theorems for integrals in the complex plane.  The approach is conceptually simple, since it is identical to what one might do by hand as part of a textbook exercise, but as an automated tool it is extremely powerful and the temporal integrals can be performed without knowing what physical problem the diagram represents.  Hence the results of AMI are analytic expressions that are valid for any problem, in any dimensionality, at arbitrary temperature.  

The initial problem of evaluating an $m$th order diagram in $D$ dimensions would involve solving $mD$ spatial integrals and $m$ temporal integrals for a total of $m(D+1)$ nested integrals.  The application of AMI resolves analytically the $m$ temporal integrals, and is easily accomplished and stored on a timescale often of micro-seconds.\cite{libami}  More importantly, the analytic result can be evaluated on the real frequency axis by replacing the external Matsubara frequency via analytic continuation $i\nu_{ext}\to \omega +i0^+$.  
The down side of this approach is that the remaining spatial integrals involve very sharp functions in a very sparse space.  One can mitigate this issue using renormalization schemes\cite{burke:2023} but nominally one remains limited by the rate of obtaining samples - the direct evaluation of an analytic function.  

It is this limitation that motivates the current implementation \texttt{torchami}.  The intent is to create a single compute framework that automates parallelization as well as allowing for easy computation using either CPUs or GPUs as the compute device.  In contrast to the existing library \texttt{libami}, we will see that the evaluation on GPUs is massively useful for generating large numbers of samples of the internal space.

\section{New Algorithms and Features}

\subsection{Algorithm: Pole Tree storage}
The construction of the AMI integrand is similar to that used in the \texttt{libami}\cite{libami} term-by-term construction scheme but with an additional factorization layer that requires an entirely different storage structure.  We introduce the \texttt{fermi\_tree} class which is a symbolic storage format for repeated sum and products of Fermi/Bose functions.  The implementation is somewhat more involved than the simple product of functions used in \texttt{libami}.  The new \texttt{fermi\_tree} structure is a natural representation that more closely mimics how these problems are solved and factorized analytically by hand.  

\begin{figure}
    \centering
    \begin{tikzpicture}[node distance=5.25cm,auto]
    \node[regular polygon,
    draw,
    regular polygon sides = 4,
    text = blue, minimum size=1cm] (L1) at (0,0) {\Huge$\times$};
    \node[regular polygon,
    draw,
    regular polygon sides = 4,
    text = black, minimum size=1cm, below left=1cm and 1cm of L1] (L2-1)  {\Huge$+$};
\node[regular polygon,
    draw,
    regular polygon sides = 4,
    text = black, minimum size=1cm, below right = 1cm and 1cm of L1] (L2-2)  {\Huge$+$};

    \node[regular polygon,
    draw,
    regular polygon sides = 4,
    text = black, minimum size=1cm, below left=1cm and .05cm  of L2-1, text width=.4cm,label=west:$+1$] (L3-1)  {$z_1$};
\node[regular polygon,
    draw,
    regular polygon sides = 4,
    text = black, minimum size=1cm, below right =1cm and .05cm of L2-1, text width=.4cm,label=west:$-1$] (L3-2) {$z_2$};

    \node[regular polygon,
    draw,
    regular polygon sides = 4,
    text = black, minimum size=1cm, below left=1cm and 0.05 cm of L2-2, text width=.4cm,label=west:$+1$] (L3-3)  {$z_3$};
\node[regular polygon,
    draw,
    regular polygon sides = 4,
    text = black, minimum size=1cm, below right=1cm and .05cm of L2-2, text width=.4cm,label=west:$+1$] (L3-4) {$z_4$};

    \path [line, line width=.8mm] (L1) -- (L2-1);
      \path [line, line width=.8mm] (L1) -- (L2-2);
      \path [line, line width=.8mm] (L2-1) -- (L3-1);
      \path [line, line width=.8mm] (L2-1) -- (L3-2);
      \path [line, line width=.8mm] (L2-2) -- (L3-3);
      \path [line, line width=.8mm] (L2-2) -- (L3-4);
    \end{tikzpicture}
    \caption{Graph representation of analytical function of a product of sums of Fermi/Bose functions.  Graph translates to $\left[f(z_1)-f(z_2) \right]\left[ f(z_3)+ f(z_4) \right]$. }
    \label{fig:treeflow}
\end{figure}
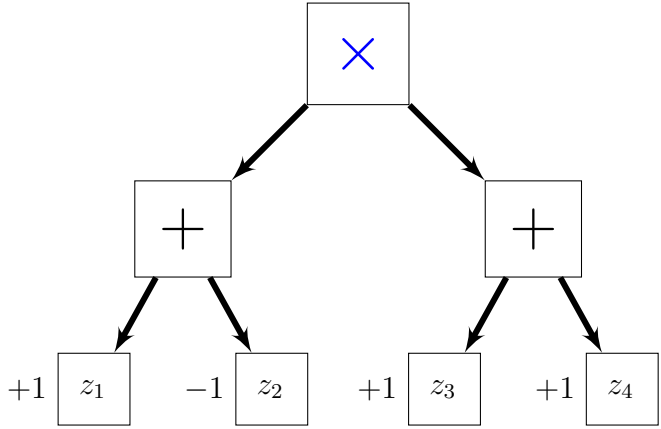

After each step of Matsubara sums, there is a term-by-term factorization that occurs.  To facilitate this we represent products and sums via a tree-graph where each node represents an operation.  The possible operations are addition, multiplication or evaluation.  In the case of evaluation, this occurs only at the bottom level of the graph, where each vertex has been assigned a pole (argument of Fermi/Bose function) and a prefactor.  An example of such a tree and its corresponding function representation is shown in Fig.~\ref{fig:treeflow}.

\subsection{Algorithm: Construction}
Following the original AMI structure\cite{AMI, libami} we begin with an equation in the form
\begin{equation}
    I^{(n)}=\frac{1}{\beta^n}\sum \limits _{\{\nu_n\}}\prod
\limits_{j=1}^N G^j(\epsilon ^j, X^j ), \label{eqn:goal}
\end{equation}
where $n$ 
is a number of summations over Matsubara frequencies $\{\nu_n\}$  and 
internal momenta $\{k_n\}$ while $N$
is the number of internal lines representing bare Green's functions $G(\epsilon,X)$.
We choose a sign convention such that the bare Green's function of the $j$th internal line is
\begin{eqnarray}
G^j(\epsilon^j, X^j) =  \frac{1}{ X^j + \epsilon^j},\label{eqn:Green_Function}
\end{eqnarray}  
where $X^j$ is the frequency and $\epsilon^j = -\epsilon(k^j)$ is the negative of the free particle dispersion, where $k^j$ is the momentum of the $j$th Green's function. 
Constraints derived from energy and momentum conservation at each vertex 
allow us to express the parameters of each free-propagator line as linear combinations of internal $\{\nu_n, {k}_n\}$ and external $\{ \nu_\gamma, k_\gamma\}$ frequencies and momenta, where $k^j  =  \sum_{\ell=1}^m \alpha_{\ell}^j k_{\ell} $, $X^j   =   \sum_{\ell=1}^{m}i\alpha_\ell^j \nu_{\ell}$.

The result of each Matsubara summation  over the fermionic frequency $\nu_p$ is given by the residue theorem,
\begin{eqnarray}\label{E: Res_Th}
\sum_{\nu_p} h(i\nu_p) = \beta \sum_{z_p} f(z_p){\rm Res}[h(z)]_{z_p},
\end{eqnarray}
where $f(z)$ is the Fermi-distribution function
and $z_p$ are the poles of $h(z)$. 
The pole of the  $j$th Green's function with respect to the frequency $\nu_p$ 
exists so long as the coefficient $\alpha_p^j$  is non-zero, and is given by 
\begin{eqnarray}\label{E: Poles_one_h(z)}
z_p ^{(j)} = -\alpha_{p} ^{j} ( -\epsilon^j + \sum_{\ell=1, \ell\neq p} ^m i\alpha_\ell^{j}\nu_\ell)  \ \ \ \ \text{for} \ \ \ \alpha_{p}^{j} \neq 0 .
\end{eqnarray}

If $h(z)$ has a pole of order $M$ at $z=z_0$, then the residue is given by 
\begin{eqnarray}\label{E: Res_mul}
Res[h(z_0)] = \frac {1} {(M-1)!} \lim_{z\to z_0} \frac {d^{M-1}}{dz^{M-1}} \bigg \{(z-z_0)^Mh(z)\bigg \}. \nonumber \\
\end{eqnarray}

The result of repeated applications of Eq.~\ref{E: Res_Th} is an analytic expression that must be stored after each Matsubara sum.  The result can be factorized into a prefactor, a \texttt{fermi\_tree}, and a product of Green's functions as in Eq.~\ref{eqn:goal} that are stored together as a struct \texttt{TamiBase::term} and vectors of such datatypes, \texttt{TamiBase::terms}.  Only the product of Green's functions needs to be carried to the next summation. Since each summation produces a number of terms, there is a factorization step that allows us to collect products and sums of Fermi/Bose functions into the \texttt{fermi\_tree} structure.  This can however be time consuming during the construction and can be disabled by setting the global flag \texttt{FACTORIZE=false}. In that case the \texttt{fermi\_tree} will only have products and this collapses down to the form of the original AMI implementation\cite{libami}, while still being in a form amenable to GPU acceleration via \texttt{pytorch}.  
A final note is that each term of the resulting integrand is in the form
\begin{equation}
    I=A\cdot FT(\{\epsilon\},\beta)\cdot R(\{\epsilon\},\{\nu\}),
\end{equation}
where $A$ is a constant prefactor, $FT$ is a fermi-tree, and $R$ is a product of Green's functions and $\beta$ is the inverse temperature.  We note that $FT$ is not a function of the set of external frequencies $\{\nu\}$.  This leads us to implement the option to evaluate a set of frequencies simultaneously which saves repeated evaluations of the $FT$ functions as well as the energies $\{\epsilon\}$. This results in potentially enormous computational savings compared with our earlier implementation.\cite{libami}

\subsection{Algorithm: Evaluation}
With the analytic solution constructed, evaluating the expressions now requires the external variables specified by the user to produce the numeric solution. From the perspective of \texttt{torchami} most of these external parameters are problem specific.  The routine only needs a set of energies appearing in the initial integrand $R0$ as well as in the external frequencies.  
This implementation allows batch evaluation of both the sets of energies as well as sets of external frequencies.  This is advantageous since the factorized \texttt{fermi\_tree} is not a function of external frequencies.  It can therefore be evaluated a single time and used as a frequency independent prefactor to be multiplied into the frequency dependent product of Green's functions.  We therefore define two datatypes to store these external properties for an $m\textsuperscript{th}$ order Feynman diagram
\begin{itemize}
    \item \texttt{energy\_t} - a rank 2 \texttt{at::Tensor} object with dimensions $N_{\textrm{E}}$ by $N$, where $N$ is the number of Green's functions in the starting integrand (Eq.~\ref{eqn:goal}) and $N_{\textrm{E}}$ is the number of energy configurations to be simultaneously evaluated (energy batch size).  Each row of the matrix contains the energies for each propagator in each configuration in the batch.
    \item \texttt{frequency\_t} - a rank 2 \texttt{at::Tensor} object with dimensions $N_{\textrm{F}}$ by $m+n_{\nu}$, where $m$ is the number of Matsubara sums and $n_{\nu}$ is the number of external frequencies. $N_{\textrm{F}}$ is the number of external frequency values that will be evaluated (frequency batch size) for each of the $N_{\textrm{E}}$ energy vectors provided in the \texttt{energy\_t} object. Each row contains the frequency vectors according to the pre-established labelling of the Feynman diagram. We use the convention that the external frequency(s) are the last elements of the rows.
\end{itemize}

Taking advantage of the parallelization provided by the \texttt{pytorch} framework, the Feynman diagram provided in the $R0$ object is evaluated for the $N_{\textrm{F}}$ external frequencies at all of the $N_{\textrm{E}}$ energies returning a rank 2 \texttt{at::Tensor} object with dimensions $N_{\textrm{F}}$ by $N_{\textrm{E}}$. Wherein each row corresponds to the Feynman diagram evaluated with the same external frequency set for various energies. Typically in a Monte Carlo integration scheme, the integrand is sampled at a various energies for a single external frequency and then repeated for all the desired frequencies. Through this new library, one is able to perform a calculation on a complete spectrum of frequencies simultaneously by summing the rows of the returned object as prescribed by a Monte Carlo integration scheme. Later we display the memory constraints on the batch sizes $N_{\textrm{F}}$ and $N_{\textrm{E}}$ which show that one evaluation may not provide a sufficient number of samples for a result within desired precision.  However one could repeatedly call this function for different energy batches and tallying the results at an optimal $N_{\textrm{E}}$, determined by the memory on the user's computing device. This method of a Monte Carlo integration takes advantage of the fact that the factorized \texttt{fermi\_tree} object does not depend on the external frequency set and therefore is evaluated once for $N_{\textrm{E}}$ energies rather than for each $N_{\textrm{E}}$ as one would do with an integration scheme using the \texttt{libami} framework.

\subsection{Feature: Pretty printing and latex output}
In addition to improved numerical stability and maximal reduction in computational expense, this new pole tree storage format also lends itself to conversion to analytic expressions via latex.  This is however an experimental feature of the library at this time.  There are a set of functions beginning with `\texttt{pretty\_print}' which produces the latex code for various components.  The function \texttt{pretty\_print\_ft\_terms} is the recommended output function. 

\subsection{Feature: Feynman Graph Representation}
To ease the use of this AMI implementation for common perturbative problems we include a simplified graph representation as well as import and labelling functions.  These allow a user to choose a diagram and quickly generate the \texttt{torchami} inputs. 

\begin{figure}
    \centering
    \includegraphics[width=\linewidth]{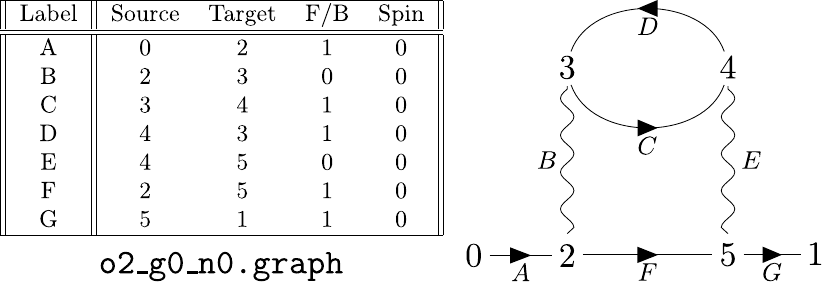}
    \caption{Example of a diagram file for a second order self energy diagram. Here each row in the graph file has been assigned a letter to emphasize its corresponding line in the Feynman diagram. Source and Target label the vertices that each line connects, F/B represents a Boolean value to indicate if the line is fermionic (1) or bosonic (0) and Spin is a integer that can be used to identify different line properties.}
    \label{fig:graph_labeling}
\end{figure}

Graphs to represent Feynman diagrams are generated using the boost graph library that is already linked for the \texttt{fermi\_tree} class.  Each diagram is stored as a file and is a plane text readable, 4 column file with a space delimeter where the columns represent (source, target, B/F, spin).  Source and target are integer indexes representing vertex ID numbers ranging from 0 to $n-1$ for an $n-$vertex graph.  B/F are zeros and ones for Bosonic and Fermionic lines respectively, and spin is an integer identifier that is typically needed to identify different line properties (spin, band index).  We show an example for a simple diagram in Fig.~\ref{fig:graph_labeling}. 

Once the user specifies the graph in a file, this is loaded into a graph using the \texttt{read\_graph} function.  In order for this to be useful the user also needs a set of momentum conserving labels.  This can be accomplished using the \texttt{label\_graph} function.  This will populate the graph edges with \texttt{alpha\_t} and \texttt{epsilon\_t} labels that can be extracted and placed the input datatype \texttt{R0\_t} using the \texttt{graph\_to\_R0} function.  

These functions are not terribly robust. They are intended as a jumping off point for a user to quickly implement simple examples and give a framework for building more complicated diagram structures. 

\subsection{Feature: C++ and python Bindings}
We provide a minimal set of bindings to the c++ classes and functions via pybind.  These include the \texttt{fermi\_tree}, and diagram graph structures as well as the \texttt{tami} class for constructing and evaluating the AMI integrands.  We provide examples for both c++ and python in the examples folder.

\section{Installation, Documentation and Tests}

\subsection{Required Libraries}
Compilation of CUDA enabled codes requires the \texttt{nvidia-toolkit}, and our codes depend on \texttt{cmake}, \texttt{boost-graph}, and \texttt{pybind11}  while \texttt{pytorch} requires python3 and contains the majority of required components.  Our recommended installation on Ubuntu can be accomplished with the following setup
\begin{bashlisting}
$ sudo apt-get install cmake libboost-dev python3 python-is-python3 nvidia-toolkit pybind11
$ pip3 install torch
\end{bashlisting}

\subsection{Compilation, Documentation and Tests}
We include an example compilation of \texttt{torchami} in the 
\begin{bashlisting}
$ mkdir build && cd build
$ cmake -DCMAKE_INSTALL_PREFIX=/path/to/desired/installDIR -DMAKE_PYTAMI=ON -DPYTHON_LIBRARY_DIR=`python -c "import site; print(site.getsitepackages()[0])"` -DPYTHON_EXECUTABLE=`which python` "-DCMAKE_PREFIX_PATH=`python -c "import torch; print(torch.utils.cmake_prefix_path)"` ..    
\end{bashlisting}

\section{Example Applications}
Examples in \texttt{c++} can be compiled via:
\begin{bashlisting}
$ export LD_LIBRARY_PATH=/where/python/environment/with/torch/is/located/lib/python3.10/site-packages/torch/lib/
$ source /where/python/environment/with/torch/is/located/bin/activate
$ cd torchami/examples
$ mkdir build && cd build
$ cmake -DPYTHON_EXECUTABLE=`which python` -DCMAKE_PREFIX_PATH=`python -c "import torch; print(torch.utils.cmake_prefix_path)"` ..
$ make 
\end{bashlisting}

We provide details of some example outputs in a later section.

\section{Examples}\label{sec:examples}
We provide a handful of specific examples, where each case the procedure is the same:
\begin{enumerate}
    \item Define integrand $R0$.
    \item Construct AMI solution.
    \item Define internal parameters and evaluate.
\end{enumerate}
For each example, we evaluate with the pole tree formulation described in this work. 

\subsection{Example 1: Second Order Self-energy}
This example is defined by the \texttt{example2()} function in the examples directory \texttt{src} files.
\subsubsection{Define Integrand}
The starting integrand for the second order self energy diagram seen in Fig.~\ref{fig:diagrams} is given by
\begin{equation}\label{eqn:r0}
    \Sigma(\nu_3,k_{ext})=\frac{1}{\nu_1-\epsilon_{k_1}}\frac{1}{\nu_2-\epsilon_{k_2}}\frac{1}{-\nu_1+\nu_2+\nu_3-\epsilon_{k_3}},
\end{equation}
where $k_i$ and $\nu_i$ is $i$\textsuperscript{th} propagator's momentum and frequency. For momentum conservation $k_3=k_{ext}-k_1+k_2$. The labeling of internal propagators for the symbolic code is arbitrary but it is essential for the final index ($i=3$) be the external line. The AMI construct function performs \texttt{NINT} nested integrals from $i=1\to$\texttt{NINT}. For this second order diagram, $\texttt{NINT}=2$.

To begin, we store the frequencies in each denominator by the set of coefficients in the overall linear combination as \texttt{alpha\_t} vectors,
\begin{cpplisting}
TamiBase::alpha_t alpha_1={1,0,0};
TamiBase::alpha_t alpha_2={0,1,0};
TamiBase::alpha_t alpha_3={-1,1,1};
\end{cpplisting}
and the energy denominators as,
\begin{cpplisting}
TamiBase::epsilon_t epsilon_1={1,0,0};
TamiBase::epsilon_t epsilon_2={0,1,0};
TamiBase::epsilon_t epsilon_3={0,0,1};
\end{cpplisting}
Here, each \texttt{epsilon\_i} is a placeholder for the negative of $\epsilon_i$.  While in this case of a $m=2$ order diagram, the geometry of the \texttt{alpha\_t} and \texttt{epsilon\_t} are the same, this typically will not be true for arbitrary diagram choices.  Specifically, the length of each \texttt{alpha\_t} is related to the order of the diagram, which in this case has length $\ell=m+1$ for an $m$th order diagram. That is, $m$ internal labels and one external label.  The length of each \texttt{epsilon\_t} is the number of Green's functions given by $2m-1$.  One may note immediately the \texttt{epsilon\_t} when listed form an $N\times N$ identity matrix, which is always the starting point for virtually any problem.  If it is known in advance that two of the energies are equivalent, that is, two or more propagators have an equal \texttt{alpha\_t}, then they can subsequently be set equal. 
We then define the starting integrand via:
\begin{cpplisting}
TamiBase::g_struct g1(epsilon_1,alpha_1);
TamiBase::g_struct g2(epsilon_2,alpha_2);
TamiBase::g_struct g3(epsilon_3,alpha_3);

TamiBase::g_prod_t R0={g1,g2,g3};
\end{cpplisting}
Each \texttt{g\_struct} has an \texttt{alpha\_t} and \texttt{epsilon\_t} that can be assigned/accessed via \texttt{g\_struct.alpha\_} and \texttt{g\_struct.eps\_}, or, alternatively assigned via the constructor as in lines $1\to3$ above. One can think of each \texttt{g\_struct} as a factor in Eq.~\ref{eqn:r0}. The resulting integrand is stored as a \texttt{g\_prod\_t} which is a vector of Green's functions representing the product in Eq.~\ref{eqn:r0}. The commutativity of multiplication implies that the ordering of the three Green's functions in \texttt{R0} will not impact the result.

\subsubsection{Construct Solution}

Once the integrand is defined, the construction requires three steps. i) Instantiate the class and define containers to store the solutions. ii) Define the integration parameters \texttt{TamiBase::ami\_parms}. iii) construct the solution. 

These can be accomplished with the following lines
\begin{cpplisting}
at::Device myDev = at::kCPU;
TamiBase ami(myDev);
TamiBase::ft_terms amiterms;

int N_INT=2;
double E_REG=0;
TamiBase::ami_parms test_amiparms(N_INT, E_REG);
ami.construct(N_INT, R0, amiterms);
\end{cpplisting}

Step (i) is performed in steps $1\to3$, (ii) $5\to7$ and (iii) in line $8$. To switch from performing the calculations on a CPU to a NVIDIA GPU, the only change in the code is replacing \texttt{at::kCPU} in line \texttt{1} with \texttt{at::kCUDA}. At this time, we have only tested this code using a CUDA enabled (Nvidia) GPU. The \texttt{TamiBase::ami\_parms} object takes an argument for a numerical regulator \texttt{E\_REG} (which is typically not used and therefore set to 0) as well as the number of internal frequencies \texttt{N\_INT} to integrate over. While the \texttt{TamiBase::ami\_parms} object is not used in the snippet, it is used in the evaluation of the integrand in subsequent steps. Line $8$ calls the \texttt{construct} method where the \texttt{TamiBase::ft\_terms} object is filled with the analytic solution for the Feynman diagram stored in the passed in \texttt{Tamibase::g\_prod\_t} object.

\subsubsection{Evaluate}
Next we will describe the evaluation of the Feynman diagram integrand. Here we populate the energy and frequency \texttt{at::tensor} objects which have a \texttt{typedef} defined to \texttt{Tamibase::energy\_t} and \texttt{Tamibase::energy\_t} respectively. Where each are implicitly \texttt{pytorch}'s complex data type \texttt{c10::complex\_double} which has also been type defined to be \texttt{TamiBase::complex\_double}. We must make sure these \texttt{at::tensor} objects are on the same device as the previously defined \texttt{TamiBase} object, \texttt{myDev}. Also, recall that for a $m\textsuperscript{th}$ order diagram the size of the energy and frequency vectors will be $m+1$. This will be defined to show the process for a general order diagram with one external frequency.

As a demonstration, we will show one way of populating the \texttt{Tamibase::energy\_t} and \texttt{Tamibase::frequency\_t} objects with respective integer batch sizes: \texttt{ebatch\_size} and \texttt{fbatch\_size}. 
\begin{cpplisting}
int diagram_order=2;
int energy_size=diagram_order+1;
int freq_size=energy_size;
int ebatch_size=1000000;
int fbatch_size=100;
\end{cpplisting}

In practice the \texttt{Tamibase::energy\_t} objects will be populated according to the free particle dispersion, $\epsilon_k$ with random momenta $k$, but for the purposes of this example we can directly generate random energies on the range $(-a, a)$, $\exists a\in \mathbb{R}$. 
\begin{cpplisting}
double a=8.0;
TamiBase::energy_t energy=2.0*a*at::rand({ebatch_size,energy_size},myDev)-a;
\end{cpplisting}

Next, we populate the \texttt{fbatch\_size} frequency vectors where each external frequency is a fermionic Matsubara frequency, $i\omega_n = \frac{(2n+1)\pi}{\beta}$ ($\beta=1$), in the last position in each vector. This is accomplished by forming a \texttt{std::vector<at::tensor>} wherein each tensor is a row vector containing the frequency vector of each propagator. Once the frequency vector is generated for all the frequency batches, the desired \texttt{frequency\_t} object is obtained by vertically stacking all the row frequency vectors, accomplished via \texttt{at::vstack}.
\begin{cpplisting}
std::vector<at::Tensor> freq_vecs = {};
for (int i=0; i < fbatch_size; ++i){
    at::Tensor row = at::full({1,freq_size}, TamiBase::complex_double(0,0), myDev);
    row[0][-1] = TamiBase::complex_double(0, M_PI*(2*i+1));
    freq_vecs.push_back(row);
}
TamiBase::frequency_t frequency=at::vstack(freq_vecs);
\end{cpplisting}

Once the \texttt{energy\_t} and \texttt{frequency\_t} objects are created, along with the inverse temperature, $\beta$, we can construct the \texttt{TamiBase::ami\_vars} object which bundles the external parameters, $\epsilon$, $\nu$, $\beta$ together. We can then call the \texttt{TamiBase::evaluate} function to obtain the resulting \texttt{at::Tensor}. This \texttt{at::Tensor} contains the integrand evaluated at the the energy vectors defined in the \texttt{energy\_t} object down the columns and the frequency vectors in the \texttt{frequency\_t} object along the rows.

\begin{cpplisting}
double BETA=1.0;
TamiBase::ami_vars external(energy, frequency,BETA);
at::Tensor result=ami.evaluate(test_amiparms, amiterms, avars);
\end{cpplisting}
In practice, following a Monte Carlo integration scheme, one would then average the rows of the \texttt{result} object to obtain the Feynman diagram as a function of the external frequencies for the pre-specified external parameters, such as $k$ and $\beta$.

\subsection{Spatial integrals}

When we want to evaluate the spatial integrals we lose the generality of the AMI process and need to specify the free dispersion $\epsilon_k$. As alluded to before, since these integrals are typically over a high-dimensional space, the most common method of integration is a Monte Carlo integration scheme, but any method can be employed. To do this we must sample the integrand at randomly chosen wavevectors $k$. In this section we will describe an example of how a Monte Carlo integration is performed with this library.

For this example we will use the python version of \texttt{torchami}, \texttt{pytami}. Code performing the following procedure can be found in the \texttt{pytami\_examples} folder of this library's github page. 

To use any external integration tool, we first need a simplified integral function. This is handled by the \texttt{AMI\_integrand} class. This class holds all of the objects needed evaluate the temporal integral such as: 
\begin{itemize}
    \item \texttt{pytami.TamiBase}
    \item \texttt{pytami.g\_prod\_t}
    \item \texttt{pytami.TamiBase.ami\_vars}
    \item \texttt{pytami.ft\_terms}
    \item \texttt{pytami.TamiBase.ami\_parms}.
\end{itemize}
 External variables such as $\beta$ and $\vec{k}$ are passed under an instance of a \texttt{ext\_vars} object. Providing am initial \texttt{frequency\_t} object with more than one frequency vector will result in integrating the Feynman diagram for all of the provided frequencies.

To make this integrand class compatible with a typical python integrator, it must be callable in python. The \texttt{AMI\_integrand} class has the python dunder method \texttt{\_\_call\_\_} defined to take an $N_{\textrm{E}}$ by $d \cdot m$ \texttt{torch.tensor} and return a $N_{\textrm{F}} \times N_{\textrm{E}}$ matrix of the integrand evaluated at each row's internal momenta configuration at all $N_{\textrm{F}}$ external frequencies. Here, $N_{\textrm{E}}$ denotes the batch size energy of the integration, $d$ is the dimensionality of the problem (2D, 3D, etc.) and $m$ is the order of the Feynman diagram. This is where the GPU acceleration is taken advantage of by evaluating a batch of $N_{\textrm{E}}$ integrands simultaneously for all $N_{\textrm{F}}$ provided frequencies.

Here we describe the process of the momenta integration more in-depth. First we must take the $N_{\textrm{E}}$ by $d \cdot m$ matrix of inputs, where the rows are the internal choices of momentum and convert them into the correct linear combinations. We will label this matrix as $k_{int}$ and look at the rows as

\begin{equation}
    k_{int} = 
    \begin{pmatrix}
        k^{(1)}_1  & k^{(2)}_1 & \hdots & k^{(d)}_1 & k^{(1)}_2 & k^{(2)}_2 & \hdots & k^{(d)}_m \\
        \vdots & \vdots & \vdots & \vdots & \vdots & \vdots & \vdots & \vdots 
    \end{pmatrix}
    \in \mathbb{R}^{N_{\textrm{E}}\times m\cdot d},
\end{equation}
where $k^{(\alpha)}_\beta$ denotes the $\alpha$ cartesian component of the the $\beta$ internal propagator of the Feynman diagram.

We can then form the complete momentum input tensor $k$ by appending copies of the external momentum $q\in \mathbb{R}^{1\times d}$
\begin{equation}
    k = 
    \begin{pmatrix}
        k_{int} & \mathbf{1}^{N_{\textrm{E}}\times 1}\cdot q
    \end{pmatrix}
    \in \mathbb{R}^{N_{\textrm{E}}\times (m+1)\cdot d},
\end{equation}
where $\mathbf{1}^{N_{\textrm{E}\times 1}}$ denotes the ones matrix of size $N_{\textrm{E}\times 1}$.

 Then we get the correct linear combinations by using the matrix of the $\alpha$ vectors from AMI which we will just call 
 \begin{equation}
     \alpha \in \mathbb{Z}^{(2m-1)\times(m+1)}.
 \end{equation}

As a recap, before with a batch size of one ($N_{\textrm{E}}=1$), we would use the alpha matrix to form the correct linear combinations as $\alpha k$. Here is an example of this with second order diagram

\begin{equation}
    \alpha = 
    \begin{pmatrix}
    1 & 0 & 0 \\
    0 & 1 & 0 \\ 
    -1 & 1 & 1 \\
    \end{pmatrix}.
\end{equation}

\noindent
Where we would use column vectors of internal and external momentum

\begin{equation}
    k = 
    \begin{pmatrix}
        k_1 \\
        k_2 \\
        q
    \end{pmatrix}.
\end{equation}

\noindent
So that the correct linear combination for this diagram is given by

\begin{equation}
    \alpha k = 
    \begin{pmatrix}
        1 & 0 & 0 \\
        0 & 1 & 0 \\ 
        -1 & 1 & 1 \\
    \end{pmatrix}
    \begin{pmatrix}
        k_1 \\
        k_2 \\
        q
    \end{pmatrix} = 
    \begin{pmatrix}
        k_1 \\
        k_2 \\
        -k_1 + k_2 + q
    \end{pmatrix}
    \label{eqn:second_ord_batchsize1_eg}
\end{equation}

\noindent
Then this matrix multiplication would be performed for each dimension.

Returning to our non-unity batch size, we do not have the internal momenta as column vectors. One solution would be to take the transpose of $k$, but since this will be a large memory object, this would be an expensive operation to perform each iteration. Alternatively, we can setup the transpose of $\alpha$ once and store it for later iterations.

Also, instead of repeatedly multiplying like in Eq.~(\ref{eqn:second_ord_batchsize1_eg}) for each spatial dimension, we can rearrange our matrix to perform this in one multiplication. By using the Kronecker tensor product with the identity matrix of size $d$, we can expand the $\alpha$ matrix to have the correct coefficients for each dimension. 

\begin{equation}
    A = \alpha \otimes \mathbb{I}_d \in \mathbb{Z}^{d(2m-1)\times d(m+1)}
\end{equation}

As an example, we can return to the second order self energy diagram in 2D $(m=2, d=2)$

\begin{equation}
    A = \alpha \otimes \mathbb{I}_2 = 
    \begin{pmatrix}
        1 & 0 & 0 \\
        0 & 1 & 0 \\ 
        -1 & 1 & 1 \\
    \end{pmatrix}
    \otimes
    \begin{pmatrix}
        1 & 0 \\
        0 & 1
    \end{pmatrix} = 
    \begin{pmatrix}
        1 & 0 & 0 & 0 & 0 & 0 \\
        0 & 1 & 0 & 0 & 0 & 0 \\
        0 & 0 & 1 & 0 & 0 & 0 \\
        0 & 0 & 0 & 1 & 0 & 0 \\
        -1 & 0 & 1 & 0 & 1 & 0 \\
        0 & -1 & 0 & 1 & 0 & 1 
    \end{pmatrix}.
\end{equation}

So we then multiply $k$ with $A^T$ to get the columns of the correct linear combinations

\begin{equation}
    \begin{split}
        k \cdot A^T
        =&\begin{pmatrix}
           k^{(1)}_1  & k^{(2)}_1 & k^{(1)}_2  & k^{(2)}_2 & q^{(1)}  & q^{(2)} \\
            \vdots & \vdots & \vdots & \vdots & \vdots & \vdots 
        \end{pmatrix}
        \begin{pmatrix}
            1 & 0 & 0 & 0 & -1 & 0 \\
            0 & 1 & 0 & 0 & 0 & -1 \\
            0 & 0 & 1 & 0 & 1 & 0 \\
            0 & 0 & 0 & 1 & 0 & 1 \\
            0 & 0 & 0 & 0 & 1 & 0 \\
            0 & 0 & 0 & 0 & 0 & 1
        \end{pmatrix}\\
        =&\begin{pmatrix}
            k^{(1)}_1  & k^{(2)}_1 & k^{(1)}_2  & k^{(2)}_2 & -k^{(1)}_1 + k^{(1)}_2 + q^{(1)}  & -k^{(2)}_1 + k^{(2)}_2 + q^{(2)} \\
            \vdots & \vdots & \vdots & \vdots & \vdots & \vdots 
        \end{pmatrix}.
    \end{split}
\end{equation}

\noindent
We denote this final matrix containing the linear combinations of momentum by $K$
\begin{equation}
    K = k \cdot A^T \in \mathbb{R}^{N_{\mathrm{E}} \times d(2m-1)}.
\end{equation}

Now all that remains is to collect the linear combination of momentum for each dimension and obtain each Green's function's dispersion. Since we used the Kronecker product, this is done by collecting $d$ columns at a time so that

\begin{equation}
    \epsilon_i = \epsilon(K_{i\cdot d}, K_{i\cdot d+1}, ..., K_{i\cdot d+d}), \forall i\in \{0, ..., 2m-2\}
\end{equation}

Now equipped with a method of incorporating the physical problem's free particle dispersion into the Monte Carlo scheme, one can compute Feynman diagrams for a desired frequency spectrum provided in a \texttt{frequency\_t} object as shown in the example \texttt{python} script \texttt{momentum\_main.py}.

\section{Benchmarks}\label{sec:benchmarks}
\begin{figure}
    \centering
    \includegraphics[width=1\linewidth]{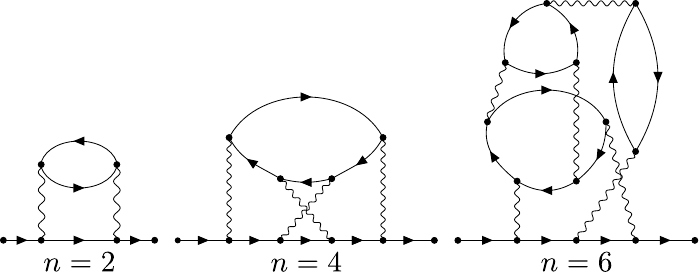}
    \caption{Feynman diagrams of orders $n = 2, 4, 6$ used to benchmark the \texttt{torchami} library.}
    \label{fig:diagrams}
\end{figure}

In this work we provide two paths to speeding up the original \texttt{libami} code library.  These are the implementation of the pole tree evaluation structure and the GPU acceleration using \texttt{pytorch}. In this section, we will refer to 2nd, 4th, and 6th  order Feynman diagrams shown in Fig.~\ref{fig:diagrams}

\subsection{GPU Scaling results}

\begin{figure}
    \centering
    \includegraphics[width=0.8\linewidth]{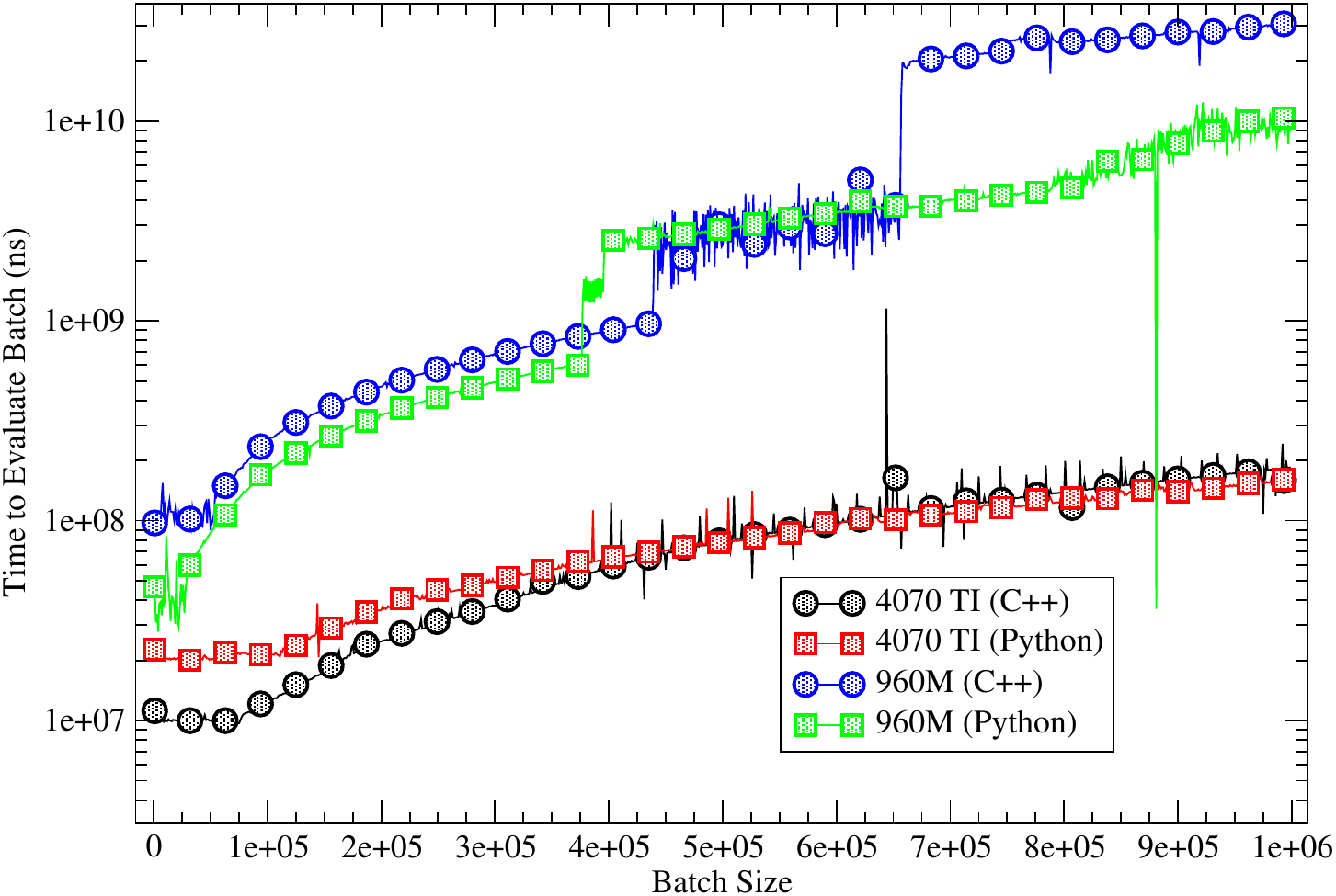}
    \caption{Time taken to simultaneously evaluate batches of a fourth order Feynman diagram as a function of the size of the batch evaluated on various Nvidia GPUs (dedicated laptop graphics, RTX 960M and a RTX 4070 TI) plotted on a logarithmic scale using both \texttt{torchami}'s native C++ implementation (circles) and Python bindings (squares). These calculations were performed with one external frequency.}
    \label{fig:log_scaling}
\end{figure}

We provide results for both the \texttt{C++} and \texttt{Python} versions of this library in Fig.~\ref{fig:log_scaling}. The scaling results shown in Fig.~\ref{fig:log_scaling} depict the time to evaluate a energy batch size as a function of the batch size with one external frequency on various GPUs. Here we see the nearly perfect scaling between the native \texttt{C++} code and \texttt{Python} bindings. In fact the \texttt{Python} bindings being faster on the older NVIDIA 960M GPU for larger batch sizes. We also see steps in the evaluation times in particular for the older architecture.  These steps are orders of magnitude and can be traced to either memory limitations or to the available cores for paralellization on the GPU. Note that on the modern graphics card, the NVIDIA 4070TI, we do not see these steps as it is able to handle much larger batch sizes. From this plot one can see how an optimal batch size can be chosen according to the hardware that the user has at their disposal in order to minimize the evaluation expense.

\subsection{Scaling with order of diagram}

\begin{figure}
    \centering
    \includegraphics[width=0.8\linewidth]{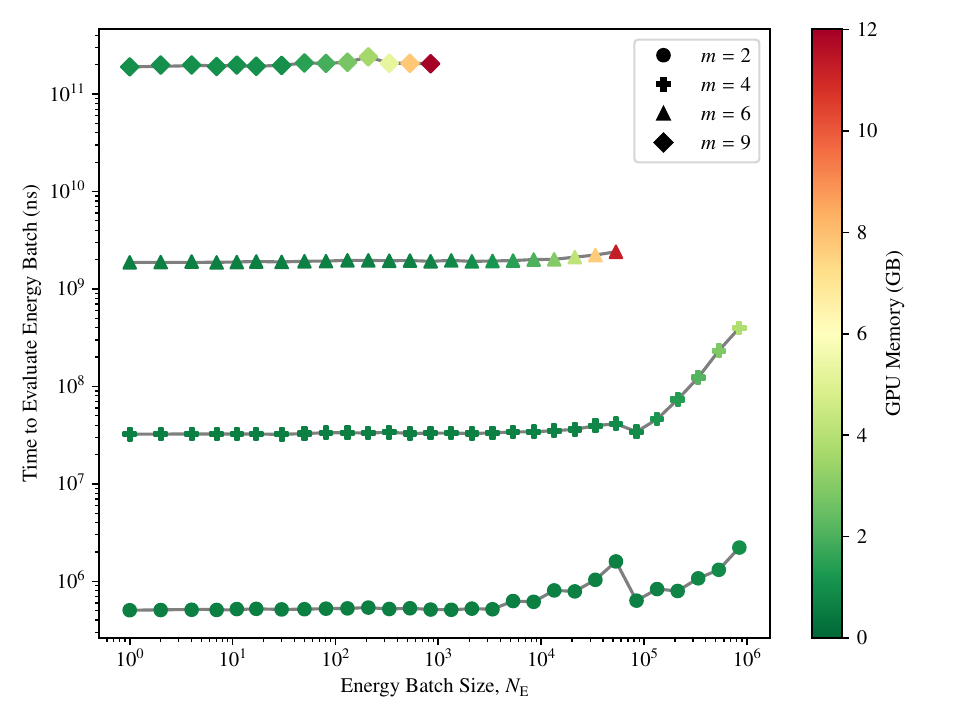}
    \caption{Time to evaluate batches integrands for diagrams of various orders, $m$, as a function of energy batch size, $N_{\textrm{E}}$. Only one frequency is computed in this data ($N_{\textrm{F}}=1$). A color map is used to show the demand on the RTX 4070Ti's 12GB of memory as a function of energy batch size. Note that for the sixth and ninth order diagram (triangles and diamonds respectively), the memory limit was exceeded before reaching a batch size of $10^6$.}
    \label{fig:RAMscaling_with_orderdiagram}
\end{figure}

Finally, in Fig.~\ref{fig:RAMscaling_with_orderdiagram}, we display how the order of the Feynman diagram, $m$, leads to different orders of magnitude in the time to evaluate a batch of energies as well as the memory usage as a function of batch size. Note that this data was collected by evaluating one frequency at a time ($N_{\textrm{F}}=1$) for various energy batch sizes, $N_\textrm{E}$. In this range of energy batch sizes, since the RTX 4070Ti card has $\approx 10^4$ cuda cores there is no appreciable cost to increasing the batch size. Crudely estimating, the time to evaluate an order $m$ diagram, $\tau_m$, is described by is an exponential scaling, $\tau_m = {10}^{0.9 m + 3.9}$. Where $\tau_m$ is measured in nanoseconds. That is, the time to evaluate a diagram increases by about an order of magnitude (${10}^{0.9}\approx10$) with each order of diagram. Also in Fig.~\ref{fig:RAMscaling_with_orderdiagram}, note that the data for orders $m=6$ and $m=9$ are shown up their respective maximum batch sizes before the 12GB of memory on the RTX 4070Ti was exceeded. Here we show that an optimal $N_{\textrm{E}}$ would need to chosen for according to the users computational device and order of diagram being evaluated. In terms of memory usage per function, the \texttt{construct} function uses about $605$MB of memory regardless of the diagram or batch size. The remaining memory usage is due to the size of the batch of energies being simultaneously evaluated.

\section{License and citation policy}
The GitHub version of \texttt{torchami} is licensed under the GNU General Public License
version 3 (GPL v. 3)\cite{gpl}. We kindly request that the present paper be cited, along with the original algorithmic paper \cite{AMI}, in any published work utilizing an application or code that uses this library.

\section{Summary}
We have presented an advanced reworking of an earlier library (\texttt{libami}\cite{libami}) for symbolic Matsubara summation and evaluation.  The tools provided in \texttt{torchami} have been expanded to include a number of important optimizations as well as both CPU and GPU support in c++ and python languages.  In addition, a minimal set of tools and examples are provided that should allow a user to create a full workflow from Feynman diagram to evaluated result for virtually any Feynman diagrammatic expansion.  Our hope is that adoption of this library (or the ideas within) will accelerate progress on real frequency evaluation of high-order perturbative expansions as has already been the case with existing works.\cite{haule:2023, kozik:2023}

\section{Acknowledgement}
We acknowledge the support of the Natural Sciences and Engineering Research Council of Canada (NSERC) RGPIN-2022-03882 and support from the Simons Collaboration on the Many Electron Problem.

\bibliographystyle{elsarticle-num}
\bibliography{refs.bib}

\end{document}